\documentclass[journal]{IEEEtran}

\usepackage{xcolor}
\usepackage[normalem]{ulem}
\usepackage{booktabs}
\usepackage{capekCommands}
\usepackage[pdftex]{graphicx}
\usepackage{cuted}
\usepackage{physics}
\usepackage{cite}
\usepackage[]{hyperref}
\hypersetup{
	colorlinks=true, 
	linkcolor=blue!70!black, 
	citecolor=red!70!black, 
	filecolor=blue!70!black, 
	urlcolor=blue!70!black, 
}

\providecommand{\Je}{\V{J}^\T{e}} 
\providecommand{\Jm}{\V{J}^\T{m}} 
\providecommand{\Ee}{\V{E}^\T{e}} 
\providecommand{\Em}{\V{E}^\T{m}} 
\providecommand{\He}{\V{H}^\T{e}} 
\providecommand{\Hm}{\V{H}^\T{m}} 

\providecommand{\Umat}{\M{U}} 
\providecommand{\Smat}{\M{U}} 

\providecommand{\Ie}{\Ivec^{\T{e}}} 
\renewcommand{\Im}{\Ivec^{\T{m}}} 

\providecommand{\Ve}{\Vvec^{\T{e}}} 
\providecommand{\Vm}{\Vvec^{\T{m}}} 

\providecommand{\Ei}{\V{E}^{\T{i}}} 
\providecommand{\Hi}{\V{H}^{\T{i}}} 

\providecommand{\fe}{\M{f}^{\T{e}}} 
\providecommand{\fm}{\M{f}^{\T{m}}} 

\providecommand{\aCircum}{r_\srcRegion} 

\graphicspath{{figures/}}
\graphicspath{{temp/}}

\begin{document}
\title{Unified Theory of Characteristic Modes:\\Part~I -- Fundamentals}
\author{Mats~Gustafsson, \IEEEmembership{Senior Member, IEEE},
Lukas~Jelinek, \\
Kurt~Schab, \IEEEmembership{Member, IEEE}, and
Miloslav~Capek, \IEEEmembership{Senior Member, IEEE}
\thanks{Manuscript received \today; revised \today. This work was supported by the  Swedish Research Council (2017-04656) and Czech Science Foundation under project~\mbox{No.~21-19025M}.}
\thanks{M. Gustafsson is with Lund University, Lund, Sweden (e-mail: mats.gustafsson@eit.lth.se).}
\thanks{L. Jelinek and M. Capek are with the Czech Technical University in Prague, Prague, Czech Republic (e-mails: \{lukas.jelinek; miloslav.capek\}@fel.cvut.cz).}
\thanks{K. Schab is with the Santa Clara University, Santa Clara, USA (e-mail: kschab@scu.edu).}
\thanks{Color versions of one or more of the figures in this paper are
available online at http://ieeexplore.ieee.org.}
\thanks{Digital Object Identifier XXX}
}

\maketitle

\begin{abstract}
A unification of characteristic mode decomposition for all method-of-moment formulations of field integral equations describing free-space scattering is derived. The work is based on an algebraic link between impedance and transition matrices, the latter of which was used in early definitions of characteristic modes and is uniquely defined for all scattering scenarios. This also makes it possible to extend the known application domain of characteristic mode decomposition to any other frequency-domain solver capable of generating transition matrices, such as finite difference or finite element methods. The formulation of characteristic modes using a transition matrix allows for the decomposition of induced currents and scattered fields from arbitrarily shaped objects, providing high numerical dynamics and increased stability, removing the issue of spurious modes, and offering good control of convergence. This first part of a two-part paper introduces the entire theory, extensively discusses its properties and offers its basic numerical validation. 
\end{abstract}

\begin{IEEEkeywords}
Antenna theory, eigenvalues and eigenfunctions, computational electromagnetics, characteristic modes, scattering, method of moments, T-matrix method.
\end{IEEEkeywords}

\IEEEpeerreviewmaketitle

\section{Introduction}
\label{Sec:Introduction}

\IEEEPARstart{C}{haracteristic} modes~\cite{Garbacz_TCMdissertation,HarringtonMautz_TheoryOfCharacteristicModesForConductingBodies} are established as a useful tool for antenna analysis and synthesis~\cite{MartaEva_TheTCMRevisited, VogelEtAl_CManalysis_PuttingPhysicsBackIntoSimulation}. The number and diversity of applications benefiting from this technique grows rapidly~\cite{ChenWang_CharacteristicModesWiley} and characteristic mode decomposition is implemented in many contemporary electromagnetic simulation tools based on integral equations. Despite this growth and widespread use, there are still unresolved theoretical issues, namely, the unique and consistent definition of the characteristic decomposition for material bodies~\cite{YlaOijala_GeneralizedTCM2019}, eigentrace tracking~\cite{SchabEtAl_EigenvalueCrossingAvoidanceInCM}, and a problematic convergence of current expansion into characteristic currents for localized sources~\cite{YeeGarbacz_SelfAndMutualAdmittancesOfWireAntennasInTermsOfCharModes, Marta_evanVid1}.

The original definition of the characteristic mode decomposition was proposed in~\cite{1948_Montgomery_Principles_of_Microwave_Circuits} and rigorously postulated in~\cite{Garbacz_TCMdissertation,GarbaczTurpin_AGeneralizedExpansionForRadiatedAndScatteredFields} for use in scattering theory.  Because this formulation is fundamentally based on scattering, it lacks many of the ambiguities and numerical issues often observed in cases when characteristic modes are used to describe antenna problems. However, this original formulation
is rarely used in contemporary work on characteristic modes due to its need for transition matrix data~\cite{Kristensson_ScatteringBook,MishchenkoTravis_TMatrixComputationsofLightScatteringbyNonsphericalParticlesReview} which is not commonly available in commercial electromagnetics solvers. Rather, it is the formulation of characteristic modes based on impedance operators~\cite{HarringtonMautz_TheoryOfCharacteristicModesForConductingBodies, Harrington_AntennaExcitationForMaximumGain} proposed later on by Harrington and Mautz that is predominantly implemented in both commercial and academic tools due to its close connection to the \ac{MoM}~\cite{Harrington_FieldComputationByMoM,HarringtonMautz_ComputationOfCharacteristicModesForConductingBodies}.  This formulation was, however, proposed phenomenologically based on its equivalence to scattering-based formulations in the case of perfectly conducting (PEC) structures \cite{HarringtonMautz_TheoryOfCharacteristicModesForConductingBodies}. Hence, as is discussed further in this work, many of the previously mentioned issues related to unambiguous definitions of modes, tracking, and convergence stem from the use of impedance, rather than scattering, operator formulations of characteristic modes. For example, the ambiguities in defining characteristic modes for material bodies arise from the fact that the nature of the impedance matrix, and its decomposition, inherently depends on what type of \ac{MoM} is utilized~\cite{Gibson_MoMinElectromagnetics, VolakisSertel_IntegralEquationMethodsForElectromagnetics, ChewTongHu_IntegralEquationMethodsForElectromagneticAndElasticWaves, Wang_GeneralizedMoM, KolundzijaDjordjevic_EMofCompositeMaterials}. This leads to multiple, sometimes conflicting, formulations; their justification, however, requires comparison with the original scattering-based definition of characteristic modes.

The question addressed in this paper is whether the scattering- and impedance-based formulations can be interlinked, \ie{}, whether the transition matrix can be expressed in terms of the impedance matrix and decomposed in a basis of outgoing waves.  This algebraic link is obtained utilizing a projection matrix~\cite{TayliEtAl_AccurateAndEfficientEvaluationofCMs} mapping basis functions representing current densities on arbitrary objects~\cite{PetersonRayMittra_ComputationalMethodsForElectromagnetics} onto a set of spherical vector waves~\cite{Kristensson_ScatteringBook}.

The proposed formulation is shown to be unique to all \ac{MoM} formulations\footnote{For example, electric field integral equation for either PEC (surface), material (volumetric) obstacles~\cite{VolakisSertel_IntegralEquationMethodsForElectromagnetics}, or PMCHWT formulation~\cite{1973_Poggio_ComputerTechniquesForElectromagnetics,1977_Wu_RS,ChangHarrington_AsurfaceFormulationForCharacteristicModesOfMaterialBodies} can be utilized.},  and, thanks to the properties of the transition matrix, many ambiguities and numerical issues are resolved. In contrast to impedance-based formulations, the transition matrix construction of characteristic modes does not exhibit issues related to spurious modes~\cite{YlaOijala_PMCHWTBasedCharacteristicModeFormulationsforMaterialBodies}. For specific geometries, the numerical dynamics of the proposed scattering-based decomposition is considerably higher than impedance-based solutions while also being considerably faster, as the problem is reduced to an ordinary eigenvalue problem of smaller size. 

Because the method examined here utilizes the transition matrix, it is not dependent on a particular numerical method~\cite{Mishchenko2020}. Rather, any method capable of delivering transition matrix data can be used. This includes both methods based on integral equations (apart from \ac{MoM}, \eg{}, discrete dipole approximation~\cite{Loke+etal2009}) and methods based on differential equations (\eg{}, finite element method~\cite{Demesy+etal2018,Fruhnert+etal2017}, finite difference frequency-domain method~\cite{Loke+etal2007}). Consequently, characteristic mode theory becomes a general frequency-domain technique which can be implemented in a variety of electromagnetic simulators, independent of a particular numerical method.

\begin{figure}[]
    \centering
    \includegraphics[width=\columnwidth]{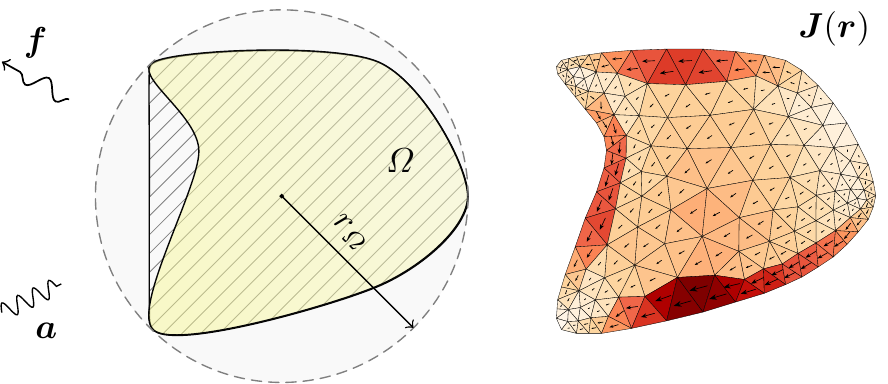}
    \caption{Left: Schematic depiction of a scatterer~$\srcRegion$ placed in the smallest circumscribing sphere of radius~$\aCircum$. Electromagnetic quantities are represented outside the circumscribing sphere in the basis of spherical vector waves with expansion coefficients collected in vectors~$\V{a}$ (regular waves) and~$\M{f}$ (outgoing waves). The characteristic modes are expressed as specific vectors~$\V{f}_n$. Hatched region represents the inner part of the convex hull of the scatterer. Right: The dominant characteristic current of the scatterer~$\srcRegion$ made of perfect electric conductor, electrical size $k\aCircum = 1$.}
    \label{fig:scat1}
\end{figure}

\section{Expansion of Electromagnetic Fields into Spherical Vector Waves}
\label{sec:ElmagDescrip}

An essential part of the proposed method is the expansion of electromagnetic fields into spherical vector waves~\cite{Stratton_ElectromagneticTheory,Kristensson_ScatteringBook}. Consider equivalent~\cite{Balanis1989} electric and magnetic current densities~$\Je$ and $\Jm$ radiating in free space\footnote{Throughout this work, \emph{free space} denotes an infinite space filled by homogeneous, isotropic, and lossless material.}~\cite{Balanis1989} confined to a scatterer~$\srcRegion$ giving rise to scattered electric and magnetic fields~$\V{E}_\T{s}$ and $\V{H}_\T{s}$, see Fig.~\ref{fig:scat1}. Assume further that we are only interested in studying fields at points outside of a sphere centered at the origin and circumscribing the support region~$\srcRegion$, \ie{}, $|\V{r}|>\aCircum$.  Under these assumptions, expansion of the scattered fields into spherical vector waves is possible via
\begin{equation}
\begin{aligned}
    \label{eq:EHExpansion}
    \V{E}_\T{s}\left(\V{r}\right) &= k\sqrt{Z}\sum_\alpha   f_{\alpha}\,\UFCN{\alpha}{4}{k\V{r}} \\
        \V{H}_\T{s}\left(\V{r}\right) &= \T{j} \dfrac{k}{\sqrt{Z}} \sum_\alpha  f_{\alpha}\,\UFCN{\overline{\alpha}}{4}{k\V{r}},
\end{aligned}
\end{equation}
where~$\UFCN{\alpha}{p}{k\V{r}}$ are spherical vector waves~\cite[Chap.~7]{Kristensson_ScatteringBook}, $k$ is the free-space wavenumber,~$Z$ is the free-space impedance and time-harmonic fields\footnote{Throughout this work, the time convention~$\exp{\T{j} \omega t}$ is assumed, where $\T{j} = \sqrt{-1}$ is the imaginary unit and~$\omega$ is the angular frequency.} are assumed.   The superindex~$p = 1$ denotes Bessel functions of the first kind (later associated with incident fields $\V{E}_\T{i}$, $\V{H}_\T{i}$ due to sources external to the region~$\srcRegion$), while~$p = 4$ denotes Hankel functions of the second kind (associated with scattered fields $\V{E}_\T{s}$, $\V{H}_\T{s}$ due to the sources $\V{J}^\T{e}$ and $\V{J}^\T{m}$ within $\srcRegion$). A bar over the index~$\alpha$ denotes a dual index which interchanges the roles of TE and TM waves for divergence-free fields~\cite[Chap.~7.2]{Kristensson_ScatteringBook}.

Because the current densities~$\Je$ and $\Jm$ radiate in free space, it is also possible to express the scattered fields $\V{E}_\T{s}$ and $\V{H}_\T{s}$ using the free-space dyadic Green's function~\cite{Zangwill_Modern_Electrodynamics}
\begin{equation}
    \V{G} \left(\V{r}_1, \V{r}_2 \right) = \left( \M{1} + \dfrac{1}{k^2} \nabla \nabla \right) \dfrac{\T{e}^{ - \T{j} k \left| \V{r}_1 - \V{r}_2 \right|} }{4\pi \left| \V{r}_1 - \V{r}_2 \right|}
\end{equation}
as
\begin{equation}
\begin{aligned}
\label{eq:EHviaGreen}
    \V{E}_\T{s} &= \Ee_\T{s} + \Em_\T{s}  = - \T{j} Z k \mathcal{L} \left\{ \Je \right\}  -   \mathcal{K} \left\{ \Jm \right\}  \\
    \V{H}_\T{s} &= \He_\T{s} + \Hm_\T{s} =  \mathcal{K} \left\{ \Je \right\}  - \T{j} \dfrac{k}{Z} \mathcal{L} \left\{ \Jm \right\}
\end{aligned}
\end{equation}
where the linear operators~$\mathcal{L}$ and $\mathcal{K}$ are defined as
\begin{equation}
\begin{aligned}
\label{eq:LK}
    \mathcal{L} \left\{ \V{J} \right\} \left( \V{r}_1 \right) &= \int \limits_{\srcRegion} \V{G} \left(\V{r}_1, \V{r}_2 \right) \cdot \V{J} \left( \V{r}_2 \right) \T{d}V_2 \\
    \mathcal{K} \left\{ \V{J} \right\} \left( \V{r}_1 \right) &= \nabla_1 \times \mathcal{L} \left\{ \V{J} \right\} \left( \V{r}_1 \right).
\end{aligned}
\end{equation}

An algebraic connection between the scattered field terms in \eqref{eq:EHExpansion} and the operator form in \eqref{eq:EHviaGreen} is given by the expansion of the dyadic Green's function into spherical vector waves~\cite{Stratton_ElectromagneticTheory,Hansen_SphericalNearFieldAntennaMeasurements, Kristensson_ScatteringBook}
\begin{equation}
    \label{eq:GreenExp}
    \V{G}\left(\V{r}_1,\V{r}_2\right)=-\T{j}k\sum_{\alpha}\M{u}_{\alpha}^{\left(4\right)}\left(k\V{r}_1\right)\M{u}_{\alpha}^{\left(1\right)}\left(k\V{r}_2\right),
\end{equation}
where, per earlier assumptions, the observation location $\V{r}_1$ is outside of the source region, \ie{}, $|\V{r}_1|>|\V{r}_2|$. 
Expanding the current densities using a suitable set of real-valued basis functions~$\{\V{\psi}_i\}$ as
\begin{equation}
    \label{eq:Jexp}
    \Je\left(\V{r}\right) \approx\sum_{i} I^\T{e}_i\V{\psi}_i\left(\V{r}\right) 
    ,\quad
    \Jm\left(\V{r}\right) \approx\sum_{i} I^\T{m}_i\V{\psi}_i\left(\V{r}\right),
\end{equation}
the relation between the formulations~\eqref{eq:EHExpansion} and~\eqref{eq:EHviaGreen} can be written as
\begin{equation}
\begin{aligned}
\label{eq:fFromIM}
\fe =  - \Umat_1 \Ie \quad \T{and} \quad \fm = \T{j} \overline{\Umat}_1 \Im,
\end{aligned}    
\end{equation}
where the matrices~$\Umat_1$ and $\overline{\Umat}_1$ are defined\footnote{In the reference~\cite{TayliEtAl_AccurateAndEfficientEvaluationofCMs} these matrices were denoted by~$\M{S}$. In this paper, symbol~$\Umat$ is used instead and the symbol~$\Smat$ is used for scattering matrices.} as
\begin{equation}
\begin{aligned}
    \label{eq:Smatrix}
    \Umat_1 = k\sqrt{Z} \left[ \, \int \limits_{\srcRegion} \M{u}_{\alpha}^{\left(1\right)} \left( k\V{r} \right) \cdot \V{\psi}_i \left( \V{r} \right) \T{d}V \right] \\
    \overline{\Umat}_1 = \dfrac{k}{\sqrt{Z}} \left[ \, \int \limits_{\srcRegion} \M{u}_{\overline{\alpha}}^{\left(1\right)} \left(k \V{r} \right) \cdot \V{\psi}_i \left( \V{r} \right) \T{d}V \right],
\end{aligned}
\end{equation}
with the interpretation of being projections of regular and real-valued spherical vector waves onto the chosen basis functions~$\{\V{\psi}_i\}$~\cite{TayliEtAl_AccurateAndEfficientEvaluationofCMs}. Analogously to relation~\eqref{eq:EHviaGreen}, the expansion coefficients~$\M{f}$ in \eqref{eq:EHExpansion} are given as the sum of contributions $\M{f}^\T{e}$ and $\M{f}^\T{m}$ from electric and magnetic current densities, respectively. 

The equivalent current densities~$\Je$ and $\Jm$ may be induced by an incident field with sources located outside of the sphere circumscribing the scatterer.  Within this circumscribing sphere, such an incident field may be expanded as
\begin{equation}
\begin{aligned}
    \label{eq:EHIExpansion}
    \V{E}_\T{i}\left(\V{r}\right) &= k\sqrt{Z}\sum_\alpha a_{\alpha}\,\UFCN{\alpha}{1}{k\V{r}} \\
        \V{H}_\T{i}\left(\V{r}\right) &= \T{j} \dfrac{k}{\sqrt{Z}} \sum_\alpha a_{\alpha}\,\UFCN{\overline{\alpha}}{1}{k\V{r}}.
\end{aligned}
\end{equation}

Within the aforementioned notation, a scattering obstacle's transition\footnote{Notice that in early texts on characteristic modes, such as~\cite{GarbaczTurpin_AGeneralizedExpansionForRadiatedAndScatteredFields,HarringtonMautz_TheoryOfCharacteristicModesForConductingBodies}, this matrix is called a perturbation operator.} matrix~$\M{T}$ relates the coefficients of regular and outgoing spherical waves as~\cite[Chap.~7.8]{Kristensson_ScatteringBook}
\begin{equation}
    \label{eq:Tmat}
    \M{f} = \M{T}\M{a}.
\end{equation}
Alternatively, decomposing regular spherical waves into outgoing and in-going waves leads to a scattering matrix 
\begin{equation}
    \label{eq:Sdef}
    \M{S} = \M{1}+2\M{T}
\end{equation} 
relating the associated sets of coefficients~\cite[Chap.~7.8]{Kristensson_ScatteringBook}, where~$\M{1}$ denotes an identity matrix.

By virtue of the orthogonality of spherical vector waves over spherical surfaces, orthogonality and radiation properties of scattered fields of the form of \eqref{eq:EHExpansion} may be written in terms of their associated expansion vector $\M{f}$.  Specifically, the cycle mean power $P_\T{rad}$, radiated by a field described by the vector~$\M{f}$, reads
\begin{equation}
\label{eq:PradFF}
    P_\T{rad} = \frac{1}{2}\M{f}^\herm \M{f}= \frac{1}{2Z} \int \limits_{4 \pi} \V{F}^*(\hat{\V{r}}) \cdot \V{F}(\hat{\V{r}}) \D{\Omega},
\end{equation}
where the superscript~${}^{\ast}$ denotes complex conjugation, the integration is carried out over the full solid angle, and the vector
\begin{equation}
    \V{F}(\hat{\V{r}}) = \lim \limits_{r \to \infty } r \T{e}^{\T{j} kr} \V{E}(\V{r})
\end{equation}
denotes the electric far field.  Here the vector~$\hat{\V{r}}$ represents a unit vector in the radial direction. Similarly, two far fields described by the vectors $\M{f}_a$ and $\M{f}_b$ are orthogonal if
\begin{equation}
\label{eq:FForth}
    \frac{1}{2}\M{f}^\herm_a \M{f}_b= \frac{1}{2Z} \int \limits_{4 \pi} \V{F}_a^*(\hat{\V{r}}) \cdot \V{F}_b(\hat{\V{r}}) \D{\Omega} = 0.
\end{equation}

\section{Characteristic Modes of Lossless Scatterers}
\label{sec:CM}

\begin{figure*}
\centering
\includegraphics[]{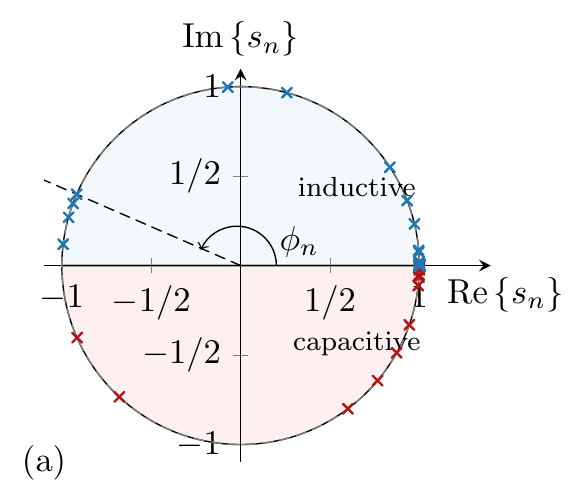}
\includegraphics[]{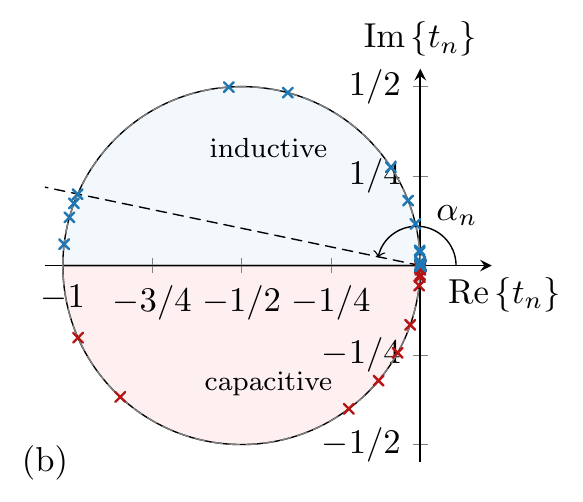}
\includegraphics[]{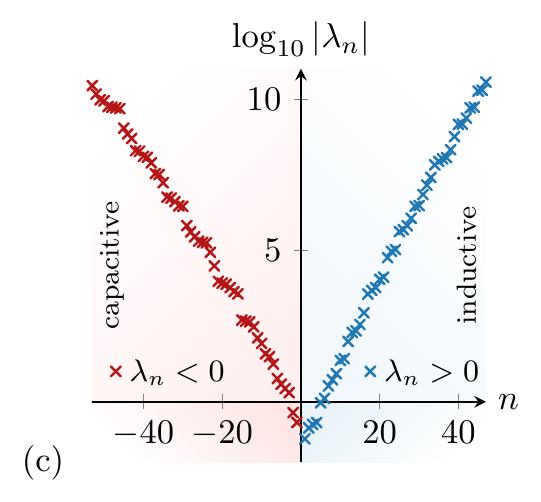}
\caption{(panel a, panel b) Eigenvalues~$s_n$ and $t_n$ as products of eigenvalue decomposition~\eqref{eq:eigS},~\eqref{eq:CM5}. (panel c) characteristic numbers~$\lambda_n$ as products of relation~\eqref{eq:tValues} or generalized eigenvalue decomposition~\eqref{eq:CM2}. All eigenvalues were evaluated for a perfectly conducting rectangular plate of aspect ratio $2:1$, and electrical size~$k\aCircum = 5$.}
\label{fig2}
\end{figure*}

\begin{table}[]
\label{tab:values}
\caption{Classification of characteristic numbers and angles depending on their physical nature.}
\centering
\begin{tabular}{cccccc} 
& $\lambda_n$ & $t_n$ & $s_n$ & $\phi_n$ & $\alpha_n$ \\ [0.5ex] 
\toprule
external res. & $0$ & $-1$ & $-1$ & $\pm\pi$ & $\pi$ \\
internal res. & $\pm \infty$ & $0$ & $1$ & $0$ & $ \pi \mp \pi/2$ \\ \midrule
capacitive & $<0$ & \multicolumn{2}{c}{$\IM\{ t_n\}, \IM\{s_n\} < 0$} & $(-\pi,0)$ & $(\pi,3\pi/2)$ \\
inductive & $>0$ & \multicolumn{2}{c}{$\IM\{ t_n\}, \IM\{s_n\} > 0$} & $(0,\pi)$ & $(\pi/2, \pi)$ \\
\bottomrule
\end{tabular}
\end{table}

Characteristic modes for lossless scatterers were originally defined as eigenvectors of scattering~\cite[Chap.~9]{1948_Montgomery_Principles_of_Microwave_Circuits} or transition (perturbation)~\cite{Garbacz_TCMdissertation} matrices.  In the latter case, characteristic modes have the interpretation of being vectors of incident or scattered field expansion coefficients which map to scaled versions of themselves through the transition matrix~\cite{Garbacz_TCMdissertation}, \ie{},
\begin{equation}
   \label{eq:CM5}
   \M{T}  \M{f}_{n} = t_n \M{f}_{n}.
\end{equation}
With $\M{f}_n$ interpreted as describing a characteristic scattered field, the corresponding characteristic incident field $\M{a}_n$ is given by \eqref{eq:Tmat} and \eqref{eq:CM5} as
\begin{equation}
    \label{eq:anToFn}
    \M{a}_n = t_n^{-1}\M{f}_n. 
\end{equation}
The eigenvalues~$t_n$ lie on a circle in a complex plane~\cite[Chap.~7]{Kristensson_ScatteringBook} and can be mapped onto the real axis by defining characteristic numbers~$\lambda_n \in \mathbb{R}$~\cite{GarbaczTurpin_AGeneralizedExpansionForRadiatedAndScatteredFields,HarringtonMautz_TheoryOfCharacteristicModesForConductingBodies} as
\begin{equation}
\label{eq:tValues}
  t_n = -\frac{1}{1 + \T{j}\lambda_n},
\end{equation}
see Fig.~\ref{fig2}. 
Characteristic numbers can be interpreted as the ratio between the imaginary and real parts of the modal complex power
\begin{equation}
P_{\T{c},n} = - \M{f}_n^\herm \M{a}_n / 2 = - t_n^{-1} P_{\T{rad},n}
\label{eq:Pc}
\end{equation}
interchanged between the modal incident field~$\M{a}_n$ and the scatterer, see Appendix~\ref{app:Sunitary}, which together with~\eqref{eq:tValues} reveals a similar identity with the eigenvalues $t_n$
\begin{equation}
	\lambda_n	=\dfrac{\T{Im} \left\{ P_{\T{c},n} \right\}}{\T{Re} \left\{ P_{\T{c},n}\right\}} = -\dfrac{\T{Im}\{t_n\}}{\Re\{t_n\}}.
\label{eq:t2la}
\end{equation} 
The eigenvalue magnitude~$|t_n|$ is commonly referred to as characteristic modal significance \cite{Austin_1998_TCM_NVIS}
\begin{equation}
   \label{eq:CM6}
  \lvert t_n\rvert= \frac{1}{\lvert 1 + \T{j}\lambda_n\rvert},
\end{equation}
as it is related to the relative significance of each characteristic mode within an expansion of a total scattered field. 

For lossless scatterers, $\M{T}$ is a normal matrix (see Appendix~\ref{app:Sunitary}) giving rise to eigenvectors~$\M{f}_n$ that are orthogonal~\cite{GarbaczTurpin_AGeneralizedExpansionForRadiatedAndScatteredFields}, may be normalized as
\begin{equation}
\M{f}_m^\herm \M{f}_n = \delta_{mn},
\label{eq:fforto}
\end{equation}
and can be chosen to be real valued for reciprocal $\M{T} = \M{T}^{\trans}$ obstacles.  By \eqref{eq:FForth}, this implies orthogonal characteristic far fields. These properties can also be observed via decomposition of the scattering matrix
\begin{equation}
\label{eq:eigS}
 \Smat \M{f}_n = s_n  \M{f}_n  = \T{e}^{\T{j} \phi_n} \M{f}_n   
\end{equation}
where\footnote{This choice is not unique and was made to obtain vanishing angle~$\phi_n = 0$ for insignificant characteristic values~$\left| \lambda_n\right| \to \infty$, $\left| t_n\right| \to 0$.}~$\phi_n = 2\arctan(\lambda_n^{-1})$ and the scattering matrix $\M{S}$ for lossless objects is unitary~\cite[Chap.~4]{Collin_FoundationsForMicrowaveEngineering}, see Appendix~\ref{app:Sunitary}. By~\eqref{eq:eigS}, the eigenvalues~$s_n$ of the scattering matrix are related to those of the transition matrix~$t_n$ by
\begin{equation}
\label{eq:sntnRelation}
s_n = 1 + 2 t_n.
\end{equation}
Graphical representation of the relations between eigenvalues~$s_n$, $t_n$ and characteristic numbers~$\lambda_n$ is given in~Fig.~\ref{fig2} while Table~\ref{tab:values} lists values and regions of physical interest, namely the scattering (external) resonances~\cite{GarbaczTurpin_AGeneralizedExpansionForRadiatedAndScatteredFields} at which the corresponding modes contribute significantly to scattering, scattering nulls (internal resonances)~\cite{GarbaczTurpin_AGeneralizedExpansionForRadiatedAndScatteredFields} at which the corresponding modes makes no contribution to scattering, and regions of capacitive/inductive modal behavior which are characterized by an excess of stored electric/magnetic energy. Table \ref{tab:values} also shows values of characteristic eigenangle~$\alpha_n$, introduced in~\cite{GarbaczTurpin_AGeneralizedExpansionForRadiatedAndScatteredFields} and later in~\cite{Newman_SmallAntennaLocationSynthesisUsingCharacteristicModes}, which is given by $\alpha_n = \pi - \arctan \left(\lambda_n \right)$ and measures the angle of eigenvalue~$t_n$, see~Fig.~\ref{fig2}.

The eigenvectors $\M{a}_n$, proportional to the vectors $\M{f}_n$ via \eqref{eq:anToFn}, can be interpreted as orthogonal incident field configurations maximizing the ratio between radiated~\eqref{eq:PradFF} and incident power
\begin{equation}
	\max_{\M{a}_n} \frac{\M{a}_n^{\herm}\M{T}^{\herm}\M{T}\M{a}_n}{\M{a}_n^{\herm}\M{a}_n}
	\quad\text{with } \M{a}_n^{\herm}\M{a}_m=0
	\quad\forall m< n,
\label{eq:RayleighQuotient}
\end{equation}  
where $\M{a}_n^{\herm}\M{a}_n$ is the incident power in the scattering matrix description~\eqref{eq:Sdef}. 
The connection between~\eqref{eq:CM5} and the Rayleigh quotient~\cite{HornJohnson_MatrixAnalysis} in~\eqref{eq:RayleighQuotient} follows from $\M{T}^{\herm}\M{T}\M{a}_n=t_n\M{T}^{\herm}\M{a}_n=|t_n|^2\M{a}_n$ for lossless cases, where vector~$\M{a}$ is real-valued, see Appendix~\ref{app:Sunitary}. This interpretation of characteristic modes could be used as an alternative physics-based definition of characteristic modes. 

In the form of~\eqref{eq:CM5}, the evaluation of characteristic modes was bound to cases where the transition matrix was known analytically~\cite{Garbacz_TCMdissertation} or cases which could be addressed via the Null-field method~\cite{Waterman1965,Kristensson_ScatteringBook}, \ie{}, mostly to spheroidal bodies. This difficulty was overcome by Harrington and Mautz~\cite{HarringtonMautz_TheoryOfCharacteristicModesForConductingBodies,HarringtonMautz_ComputationOfCharacteristicModesForConductingBodies} who used \ac{MoM} formulations of field integral equations to express characteristic modes for bodies of arbitrary shape. In their formulation, characteristic modes of perfectly conducting obstacles described by the electric field integral equation (EFIE) were eigensolutions to a generalized eigenvalue problem
\begin{equation}
    \label{eq:CM2}
    \M{Z}^\T{ee}\Ie_n = \left(1 + \T{j}\lambda_n \right) \T{Re} \left\{ \M{Z}^\T{ee} \right\} \Ie_n,
\end{equation}
where $\M{Z}^\T{ee}$ is the system matrix of the discretized electric field integral equation~\cite{Harrington_FieldComputationByMoM}
\begin{equation}
    \label{eq:MoM}
    \M{Z}^\T{ee} \Ie = \Ve,
\end{equation}
with~$\Ve$ containing the incident electric field expansion coefficients (defined below in~\eqref{eq:V}).  Through the relation 
\begin{equation}
P_{\T{c},n} = \frac{1}{2}\left(\Ie_n\right)^\herm \Ve_n  = \left(1+ \T{j} \lambda_n \right)P_{\T{rad},n},
\end{equation}
the formulation in~\eqref{eq:CM2} also clearly presents the connection between the characteristic value~$\lambda_n$ and the power ratio in~\eqref{eq:t2la}. This formulation of characteristic modes was later extended to lossless penetrable scatterers using volume~\cite{HarringtonMautzChang_CharacteristicModesForDielectricAndMagneticBodies} and surface~\cite{ChangHarrington_AsurfaceFormulationForCharacteristicModesOfMaterialBodies} formulations. 

The formulation of characteristic modes based on field integral equations brings computational flexibility, nevertheless it also allows for spurious (and unwanted) solutions~\cite{Huang+etal2019} which are not present in the original formulation~\eqref{eq:CM5}. Different possibilities in formulating field integral equations~\cite{2016_Hu_TAP} also make it difficult to identify the radiation operator that should be put on the RHS of~\eqref{eq:CM2} in order to maintain far-field orthogonality. In the following section, a proposal is made to remove these issues, keeping the computational capabilities of field integral equations, while also returning to the original definition of characteristic modes in~\eqref{eq:CM5}.

\section{Evaluation of Characteristic Modes Using Transition Matrix}
\label{sec:CMfromT}

Here we formulate a connection between the method-of-moments impedance matrix and the transition matrix $\M{T}$.  The goal of this connection is to establish a procedure for evaluating characteristic modes using the transition matrix formulation in \eqref{eq:CM5} while maintaining the flexibility afforded by impedance-based formulations, such as \eqref{eq:CM2}, particularly regarding the analysis of arbitrarily shaped obstacles. The procedure is presented on a \ac{MoM} formulation of field integral equations for penetrable bodies based on the surface equivalence principle (\eg{}, PMCHWT, Poggio-Miller-Chang-Harrington-Wu-Tsai and M\"{u}ller formulations)~\cite{1973_Poggio_ComputerTechniquesForElectromagnetics,1977_Wu_RS,ChangHarrington_AsurfaceFormulationForCharacteristicModesOfMaterialBodies,Jin_TheoryAndComputationOfElectromagneticFields}. Adaptations to other formulations are shown in~Appendix~\ref{App:IEformulations}. 

In the these formulations, the \ac{MoM} system describing the scattering scenario in Section~\ref{sec:ElmagDescrip} reads~\cite{ChangHarrington_AsurfaceFormulationForCharacteristicModesOfMaterialBodies}
\begin{equation}
\label{eq:ZIVSurfEqv}
\mqty[
\M{Z}^\T{ee}& \T{j} \M{Z}^\T{em}\\
\T{j} \M{Z}^\T{me}&\M{Z}^\T{mm}
]
\mqty[\Ie\\\T{j}\Im] = 
\mqty[\Ve\\ \T{j} \Vm],
\end{equation}
where the excitation vectors~$\Ve$ and $\Vm$ represent the incident electric field~$\Ei$ and incident magnetic field~$\Hi$ via
\begin{equation}
\begin{aligned}
    \label{eq:V}
    V^\T{e}_n &= \int \limits_{\srcRegion} \Ei\left( \V{r} \right)  \cdot \V{\psi}_n \left( \V{r} \right) \T{d}S \\
    V^\T{m}_n &= \int \limits_{\srcRegion} \Hi\left( \V{r} \right) \cdot \V{\psi}_n \left( \V{r} \right) \T{d}S,
\end{aligned}
\end{equation}
respectively, and different components of the partitioned system matrix refer to inner products between basis functions and their projections through the~$\mathcal{K},\mathcal{L}$ operators~\eqref{eq:LK} in the interior and exterior regions, see Appendix~\ref{App:IEformulations}. For brevity, the partitioned system~\eqref{eq:ZIVSurfEqv} is written as
\begin{equation}
\label{eq:ZIVsimple}
\M{Z} \M{I} = \M{V},
\end{equation}
where it is assumed that the impedance matrix~$\M{Z}$ is invertible.

Employing the field expansion~\eqref{eq:EHIExpansion}, and assuming that sources of the incident field are external to the convex hull of the scatterer~\cite{Mishchenko+etal2000} (see Fig.~\ref{fig:scat1} and related discussion in Section~\ref{sec:disc:conv}), it is  possible to write the incident fields in terms of the regular wave expansion coefficients in \eqref{eq:EHIExpansion} as
\begin{equation}
\label{eq:VWvecSW}
\M{V} = \Umat^\T{T} \M{a},
\end{equation}
where
\begin{equation}
\label{eq:UmatSimple}
\Umat = \mqty[\Umat_1 & - \overline{\Umat}_1 ].
\end{equation}
Additionally, by \eqref{eq:fFromIM}, the radiated fields produced by the electric and magnetic currents may be expressed in terms of the outgoing spherical wave expansion coefficients in \eqref{eq:EHExpansion} as
\begin{equation}
\label{eq:fTot}
\M{f} = - \Umat \M{I}.
\end{equation}

Left multiplying~\eqref{eq:ZIVsimple} by $-\Umat\M{Z}^{-1}$,  and substituting from~\eqref{eq:VWvecSW} and~\eqref{eq:fTot} results in
\begin{equation}
\label{eq:IZV2}
\M{T} = - \Umat
\M{Z}^{-1} 
\Umat^\T{T},
\end{equation}
which in the simpler form with only $\M{Z}^{\T{ee}}$ was obtained in~\cite{1991_Gurel_APS,2013_Kim_TAP,2017_Markkanen_JQSRT,Losenicky_etal_MoMandThybrid}, see also~\cite{Nieminen+etal2003,Reid2000}. When the above expression is used to express the transition matrix, the eigenvalue problem in~\eqref{eq:CM5} yields a formulation of characteristic modes with both the computational advantages of the impedance-based  description together with the robust and original transition matrix definition. Relations analogous to~\eqref{eq:IZV2} for other common field-integral formulations\footnote{Conversion of an impedance matrix~$\M{Z}$ into a transition matrix~$\M{T}$ also gives a possibility to verify the correctness of the underlying impedance matrix formulation, since the matrix~$\M{T}$ is unique to a given scatterer and has eigenvalues lying on a circle in the complex plane, see~Fig.\ref{fig2}.} are given in Appendix~\ref{App:IEformulations}.

An apparent drawback of the aforementioned formulation involving a combination of \eqref{eq:CM5} and \eqref{eq:IZV2} is that characteristic eigenvectors are given in terms of the scattered field expansion coefficients~$\M{f}$. This can be resolved by realizing that, by \eqref{eq:anToFn}, an eigenvector~$\M{f}_n$ corresponds to a particular excitation~$\M{a}_n$.
An analogous relation holds for the current eigenvectors~$\M{I}_n$, which are linked to specific excitations through~\eqref{eq:ZIVsimple} as
\begin{equation}
    \M{I}_n=\M{Z}^{-1}\M{V}_n
    =\M{Z}^{-1}\M{U}^{\T{T}}\M{a}_n
    =t_n^{-1}\M{Z}^{-1}\M{U}^{\T{T}}\M{f}_n,
    \label{eq:CMTJ}
\end{equation}
where~\eqref{eq:VWvecSW} has been used to relate the excitation vector~$\M{V}_n$ with the excitation vector~$\M{a}_n$. Therefore, even when characteristic currents are desired, there is no need to solve a generalized eigenvalue problem in a form similar to~\eqref{eq:CM2}. Rather, solving the simpler ordinary eigenvalue problem~\eqref{eq:CM5} is sufficient. It is also worth noting that the computation of characteristic currents~$\M{I}_n$ is carried out by multiplication of eigenvectors~$\M{f}_n$ with the matrix~$\M{Z}^{-1}\M{U}^{\T{T}}$ which was already used to determine the transition matrix~$\M{T}$ through relation~\eqref{eq:IZV2}. The evaluation of characteristic currents therefore requires only minor additional computational burden.  Relation~\eqref{eq:CMTJ} shows that characteristic currents are undefined at scattering nulls characterized by vanishing eigenvalue~$t_n = 0$ and infinitely large characteristic number~$\left| \lambda_n \right| \to \infty$.

\section{Evaluation of Characteristic Modes Using Currents}\label{S:CMI}

From \eqref{eq:CMTJ},  a generalized eigenvalue problem directly in characteristic currents may be formed which illustrates similarities and differences between the impedance-based~\eqref{eq:CM2} and scattering-based~\eqref{eq:CM5} methods for computing characteristic modes. Left multiplying~\eqref{eq:CMTJ} by the matrix $\M{Z}$ and substituting~\eqref{eq:fTot} for the eigenvector~$\M{f}_n$ produces the generalized eigenvalue problem 
\begin{equation}
    \label{eq:CMI}
    \M{Z}\M{I}_n = (1 + \T{j}\lambda_n) \M{R}_0\M{I}_n,
\end{equation}
where, by using $\M{U}^\herm=\M{U}^\trans$ for real-valued basis functions and real-valued regular spherical waves in \eqref{eq:Smatrix}, the radiation operator $\M{R}_0$ is defined by
\begin{equation}
\label{eq:R0}
    \M{R}_0 = \Umat^\herm \Umat.
\end{equation}
This last association is possible due to the chosen expansion into spherical waves which allows the radiated power~\eqref{eq:PradFF} to be written as
\begin{equation}
    P_\T{rad} = \frac{1}{2} \M{f}^\herm \M{f}
    =\frac{1}{2} \M{I}^\herm \Umat^\herm\Umat \M{I},
\end{equation}
see Appendix~\ref{app:Sunitary}.  The use of \eqref{eq:R0} to define the radiation operator in the eigenvalue problem \eqref{eq:CMI} is distinct from purely impedance-based characteristic mode formulations (\eg{} \cite{HarringtonMautz_ComputationOfCharacteristicModesForConductingBodies}) in that it uniquely defines the radiation matrix~$\M{R}_0$ independently of the impedance matrix~$\M{Z}$. Hence, irrespective of the particular form of the impedance matrix~$\M{Z}$, the characteristic currents in~\eqref{eq:CMTJ} computed by the problem \eqref{eq:CM5} are orthogonal with respect to matrix~$\M{R}_0$ for lossless scenarios and may be normalized as
\begin{equation}
\label{eq:IRI}
\M{I}_m^\herm \M{R}_0 \M{I}_n = \M{f}_m^\herm \M{f}_n = \delta_{mn},
\end{equation}
where the last equality stems from properties discussed in Section~\ref{sec:CM}. By virtue of \eqref{eq:FForth}, the above expression implies orthogonal characteristic far fields.  For lossless systems, relation~\eqref{eq:CMTJ}  produces characteristic currents orthogonal with respect to the impedance matrix~$\M{Z}$, \ie{},
\begin{equation}
\label{eq:IZI}
\M{I}_m^\herm \M{Z} \M{I}_n = (1 + \T{j}\lambda_n) \delta_{mn}.
\end{equation}


\section{Numerical Validation}

This section presents the basic capabilities of the aforementioned methodology on an example of dielectric sphere mostly aiming at validation and presentation of the achieved numerical precision.

\begin{figure}[t]
    \centering
    \includegraphics[width=\columnwidth]{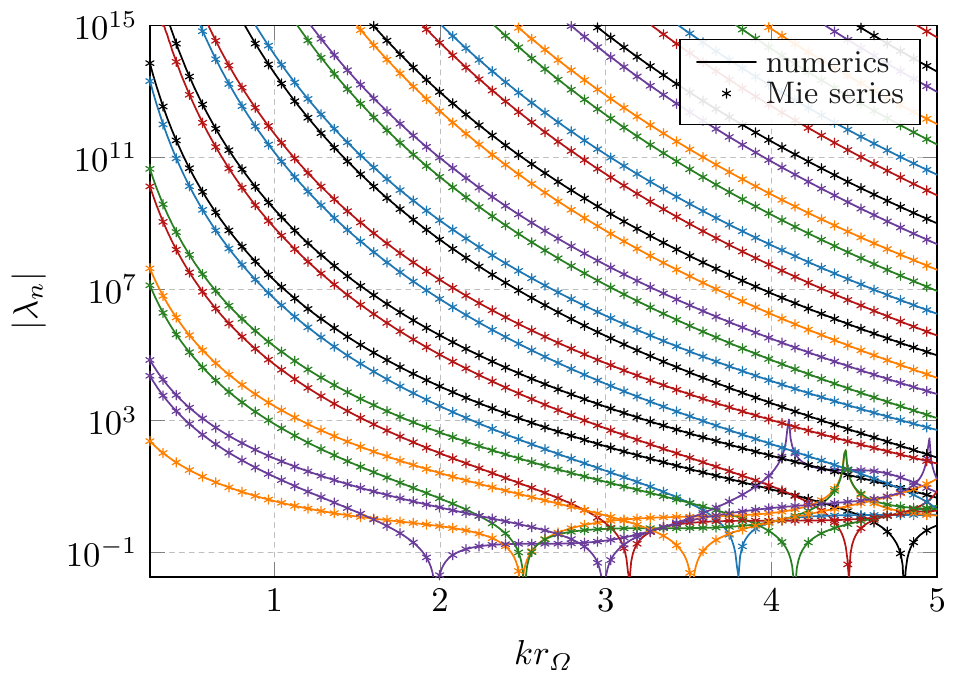}
    \caption{Magnitude of characteristic values for a dielectric sphere with relative permittivity $\varepsilon_{\mathrm{r}}=3$ calculated using PMCHWT formulation (curves) and with analytic expressions based on Mie series (markers).}
    \label{fig:dielsphere3}
\end{figure}

Characteristic numbers for a dielectric sphere with relative permittivity $\varepsilon_{\mathrm{r}}=3$ are depicted in Fig.~\ref{fig:dielsphere3}. Characteristic numbers are calculated using~\eqref{eq:CM5} and~\eqref{eq:tValues} in which the matrix~$\M{T}$ was evaluated by~\eqref{eq:IZV2} with the underlying impedance matrix taken from the PMCHWT formulation (solid curves) or in which matrix~$\M{T}$ was evaluated analytically using the Mie series~\cite{Kristensson_ScatteringBook} (markers). Colors are used to distinguish different modes. A technique used to track modes is detailed in Part~II~\cite{Gustafsson+etal_CMT2_2021}. The maximum necessary degree of spherical vector waves is estimated from~\cite{JimingSong2001}
\begin{equation}
    \label{eq:Lmax}
    L_\T{max} = \lceil k\aCircum+\iota\sqrt[3]{k\aCircum}+3 \rceil,
\end{equation}
where the parameter $\iota$ controls the accuracy. High accuracy is needed to compute characteristic numbers with large amplitude and $\iota=7$ was used in Fig.~\ref{fig:dielsphere3} with parameter~$k\aCircum=5$ for the entire frequency range. This gives the maximal degree~$L_\T{max}=20$ and leads to a total of $2L_\T{max}(L_\T{max}+2)=880$ spherical waves, however rotational symmetry reduces the number of independent waves to $2L_\T{max}=40$ using a body of revolution (BoR) formulation~\cite{Mautz_RadiationandScatteringFromBodiesofRevolution}.    

\begin{figure}[]
    \centering
    \includegraphics[width=0.49\columnwidth]{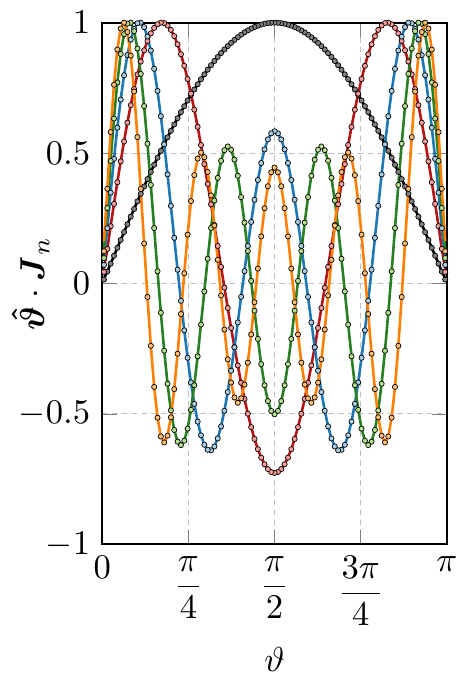}
    \includegraphics[width=0.49\columnwidth]{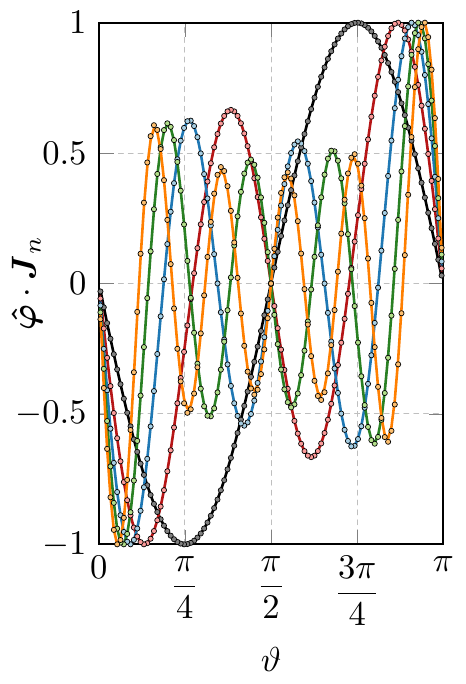}
    \caption{Profiles of characteristic currents for the same setup as in Fig.~\ref{fig:dielsphere3} and electrical size~$k\aCircum=1$. Solid curves correspond to numerical procedure~\eqref{eq:CMTJ}, while markers show analytical result. Only  modes independent on the azimuthal angle~$\varphi$ are shown and every second mode (even and odd) of the $20$~first in Fig.~\ref{fig:dielsphere3} are shown for $\vartheta$ and $\varphi$ components. For presentation purposes, peak values of modal profiles were all normalized to unity instead of using normalization~\eqref{eq:IRI}.}
    \label{fig:dielsphere}
\end{figure}

The results show that numerically evaluated characteristic numbers agree with the analytic results over a high dynamic range~\cite{TayliEtAl_AccurateAndEfficientEvaluationofCMs}. A comparison of current profiles numerically obtained from~\eqref{eq:CMTJ} and current profiles known analytically is shown in Fig.~\ref{fig:dielsphere} for the electrical size~$k\aCircum=1$. Only the dependence on the polar angle~$\vartheta$ of modes independent on the azimuthal angle~$\varphi$ is shown for brevity. The results show that the numerically computed modes agree with the analytical modes for all considered cases. 

\section{Discussion}
\label{sec:disc}

This section discusses several important points raised throughout this paper, with emphasis placed on their interpretation within the context of contemporary characteristic mode theory.

\subsection{Uniqueness of Characteristic Mode Decomposition}
\label{sec:disc:uniq}

The set of characteristic modes for a given scatterer should be unique and independent of the numerical method used to obtain them. This assumption is fulfilled with the scattering formulation~\eqref{eq:CM5} and the link between an impedance and the scattering formulation~\eqref{eq:IZV2}, which is valid for all method-of-moments formulations. Other numerical methods may also be used, \eg{}, finite element method~\cite{Demesy+etal2018,Fruhnert+etal2017}, finite difference frequency-domain method~\cite{Loke+etal2007}, or null-field method~\cite{Mishchenko2020}; so long as a transition matrix~$\M{T}$ can be constructed. This indicates that characteristic mode analysis does not require the use of the \ac{MoM}~\cite{SarkarMokoleSalazarPalma_AnExposeOnInternalResonancesCM}. The evaluation of characteristic modes using the finite element method is shown in Part~II~\cite{Gustafsson+etal_CMT2_2021}.

Another advantage of the scattering formulation is its lack of spurious modes. Once the transition matrix~$\M{T}$ is calculated, its decomposition is immune to the presence of internal resonances ($t_n = 0$). However, it should be noted that its construction via~\eqref{eq:IZV2} is difficult in this case due to potential singularities of the impedance matrix~$\Zmat$.

\subsection{Numerical Dynamics and Computational Cost}
\label{sec:disc:dyn}

Using fixed numerical precision, the scattering- and impedance-based approaches perform similarly in terms of computational time for electrically small problems. For electrically larger problems, the scattering approach is faster than a \ac{MoM} implementation of the impedance formulation. With respect to electrical size, the impedance matrix $\M{Z}$ scales quadratically for surface formulations and cubically for volume formulations, whereas the transition matrix $\M{T}$ scales quadratically in both cases.
Furthermore, the number of spherical waves is typically orders of magnitude smaller than the number of local basis functions. Hence, the transition matrix method is expected to generally require less computational time as it is its construction, not its eigenvalue decomposition, which accounts for the majority of the overall computational cost.

\begin{figure}[]
    \centering
    \includegraphics[width=\columnwidth]{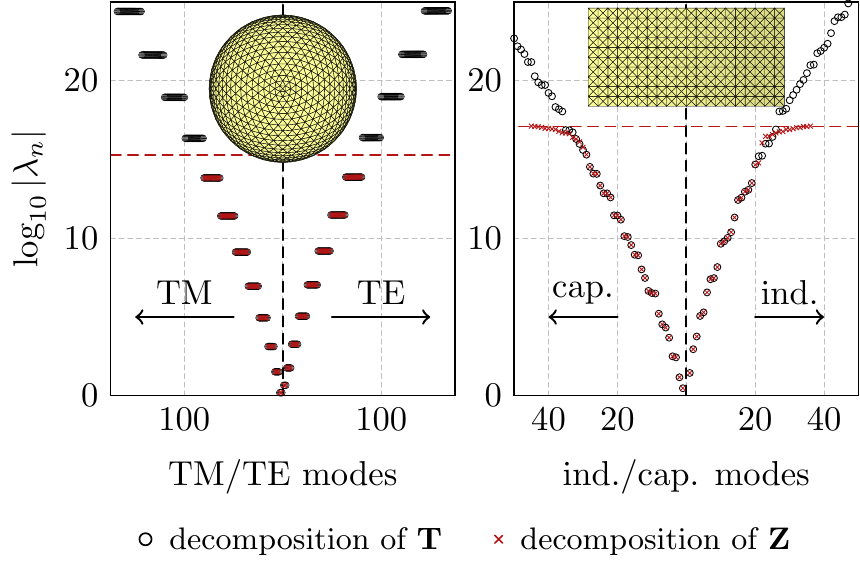}
    \caption{Characteristic modes evaluated with scattering and impedance formulation, \eqref{eq:CM5} and \eqref{eq:CMI}, respectively. A spherical shell discretized into $2400$~triangles ($3600$~RWG basis functions) and a rectangular plate of aspect ratio $2:1$ discretized into $800$~triangles ($1170$~RWG basis functions) are studied. Both obstacles are made of PEC and of electrical size~$k\aCircum = 1$. The decomposition is evaluated with generalized Schur decomposition. The characteristic numbers of the spherical shell exceeding dashed line were evaluated as infinite by the used routine and are therefore not shown.}
    \label{fig:dynamics}
\end{figure}

Figure~\ref{fig:dynamics} compares magnitudes of characteristic numbers produced by both decomposition methods applied to a spherical shell and a rectangular plate.  In the case of the spherical shell, $3600$~RWG basis functions and $510$~spherical waves ($L_\T{max} = 15$) are used in the impedance- and scattering-based methods, respectively. Despite involving significantly more unknowns, the impedance method has numerical dynamics saturated at double precision (red dashed line), while the scattering formulation might offer higher dynamics. In the case of the irregular (in the context of spherical symmetry) rectangular plate, $1170$~RWG and $510$~spherical waves are used. Here the relative size, and thus computational cost, of the two solutions are comparable, however the scattering-based method still affords improved numerical dynamics beyond those of the impedance-based solution.

Beyond computational cost and numerical dynamics, a notable advantage of scattering method over impedance method is applicability of a tracking procedure described in Part~II~\cite{Gustafsson+etal_CMT2_2021}.

\subsection{Convergence of Characteristic Mode Superposition}
\label{sec:disc:conv}

\begin{figure}[]
    \centering
    \includegraphics[width=\columnwidth]{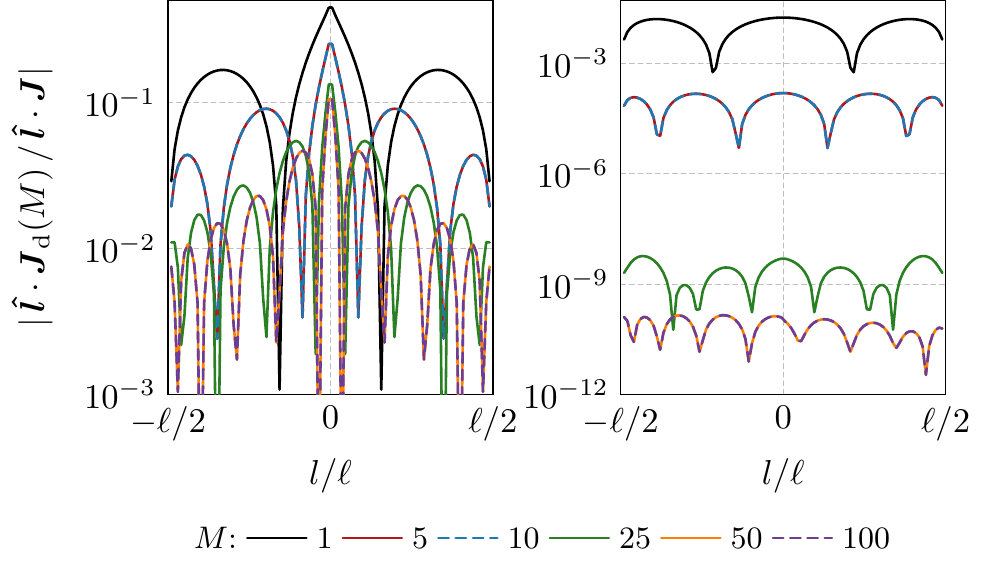}
    \caption{Relative error of the sum~\eqref{eq:IsumCM1} of characteristic modes evaluated via~\eqref{eq:CM5} and~\eqref{eq:CMTJ}. A thin-strip PEC dipole oriented along direction~$\hat{\V{l}}$, discretized into $400$~triangles ($499$~RWG basis functions), and having electrical size~$k\aCircum = 2\pi/3$ is studied. The excitation used is a delta gap feeder placed in the middle of the dipole (left) and a plane wave incident perpendicularly to the plane of the dipole and with polarization along the dipole.}
    \label{fig:convergence}
\end{figure}

Characteristic modes are known to be problematic~\cite{YeeGarbacz_SelfAndMutualAdmittancesOfWireAntennasInTermsOfCharModes,Marta_evanVid1} in describing certain scenarios, such as those in which the incident field (described by the vector~$\M{V}$) is produced by localized sources within $\srcRegion$. This issue is illustrated in Fig.~\ref{fig:convergence} using a thin-strip PEC dipole excited by a delta gap source (source within the convex hull of region~$\varOmega$) or a plane wave (far-field source). In both cases, the residual~\cite{Marta_evanVid1} current~$\Ivec_\T{d} (M) = \Ivec - \Ivec (M)$ is evaluated, where the vector~$\Ivec = \M{Z}^{-1} \M{V} $ comes from a direct solution of the scattering problem, while vector $\Ivec (M)$ is evaluated using a summation formula\footnote{Notice that the formula assumes a symmetric transition matrix (reciprocity) and that matrix transposition inside the formula can be interchanged with Hermitian conjugation in lossless scenarios where eigenvectors~$\M{f}_n, \M{I}_n$ can be chosen real-valued, see Section~\ref{sec:CM}. Unlike in~\cite{HarringtonMautz_TheoryOfCharacteristicModesForConductingBodies} we have also chosen to write the term~$\M{f}_n^\trans \M{f}_n$ in the denominator instead of usual term ~$\M{I}_m^\trans \M{R}_0 \M{I}_n$ which is only valid in specific formulations of field integral equations.}~\cite{HarringtonMautz_TheoryOfCharacteristicModesForConductingBodies}
\begin{equation}
\Ivec (M) = \sum\limits_{n=1}^M \dfrac{1}{\left( 1 + \J \lambda_n \right)} \dfrac{\Ivec_n^\trans \Vvec}{\M{f}_n^\trans \M{f}_n} \Ivec_n = - \sum\limits_{n=1}^M t_n \dfrac{\Ivec_n^\trans \Vvec}{\M{f}_n^\trans \M{f}_n} \Ivec_n.
\label{eq:IsumCM1}
\end{equation}
It is observed in Fig.~\ref{fig:convergence} that decomposition of the current~\eqref{eq:IsumCM1} excited by a delta-gap source fails to converge. 

The scattering-based description of characteristic modes shown in this paper offers an explanation to this phenomenon. The connection between transition and impedance matrix based characteristic modes is provided via relation~\eqref{eq:VWvecSW}. This relation is, nevertheless, only valid for sources which produce incident fields of the form of \eqref{eq:EHIExpansion}, \ie{}, if a sphere centered at origin exists that separates sources of incident field from the scatterer. Since the scatterer might be arbitrarily shifted with respect to origin, relation~\eqref{eq:VWvecSW} allows for the modeling of sources anywhere outside the convex hull of region~$\varOmega$, see Fig.~\ref{fig:scat1}. No such shift, however, allows for a separating sphere to be drawn for sources inside this convex hull (such as delta gap source). It should also be realized that as the source of the incident field approaches the convex hull, the radius of the corresponding separating sphere differs greatly from that of the smallest sphere circumscribing the scatterer and it is the radius of the former that should be used in~\eqref{eq:Lmax} to describe such scenario, making the summation formula~\eqref{eq:IsumCM1} impractical or even insufficient when the finite precision of numerical characteristic modes decompositions is considered.

The aforementioned issues can also be viewed from a different perspective. Imagine that the true current~$\M{I}$ induced on a scatterer is known and with it the expansion vector~$\M{f}$, see~\eqref{eq:fTot}, representing the electric far field. Employing relation~\eqref{eq:CMTJ} it can be seen that the summation formula~\eqref{eq:IsumCM1} can be rewritten as
\begin{equation}
\Ivec (M) = \sum\limits_{n=1}^M \dfrac{\M{f}_n^\trans \M{f}}{\M{f}_n^\trans \M{f}_n} \Ivec_n,
\end{equation}
where it is observed that the summation coefficients are projections of the total electric far field into modal far fields, \ie{}, a quantity that is insensitive to any near-field distributions. Characteristic modes are therefore not well suited to expand current densities on scatterers excited by sources located in the near field of the region~$\varOmega$.

Despite these difficulties, characteristic modes are appropriate to expand scattering properties, such as radiated far fields described by the vector~$\M{f}$. To that point, a formula analogous to~\eqref{eq:IsumCM1} can be written as
\begin{equation}
\M{f} (M)  =  \sum\limits_{n=1}^M t_n \dfrac{\M{f}_n^\trans \M{a}}{\M{f}_n^\trans \M{f}_n} \M{f}_n,
\label{eq:fsum}
\end{equation}
the convergence of which, unlike of~\eqref{eq:IsumCM1}, is favorable, especially in the case of electrically small scatterers.

\subsection{Spherical Waves as Extended Ports}
\label{sec:disc:unif}

An interesting point of view on the connection between transition and impedance matrices~\eqref{eq:IZV2} is obtained through examination of a network representation~\cite{Capeketal_OptimalityOfTARCAndRealizedGainForMultiPortAntennas} and port modes~\cite{HarringtonMautz_ControlOfRadarScatteringByReactiveLoading}. The port admittance matrix~$\M{y}$ of a multiport antenna is expressed as
\begin{equation}
\M{y} = \M{P}^\trans \M{Z}^{-1} \M{P} = \M{P}^\trans \M{Y} \M{P},
\label{eq:pp1}
\end{equation}
where the columns of the matrix~$\M{P}$ define ports~\cite{Capeketal_OptimalityOfTARCAndRealizedGainForMultiPortAntennas}, \ie{}, the excitation vector~$\M{V}$ is given as
\begin{equation}
\M{V} = \M{P} \M{v},
\label{eq:pp3}
\end{equation}
where~$\M{v}$ is a column vector of port voltages. Relation~\eqref{eq:pp1} resembles~\eqref{eq:IZV2} provided that matrix~$\M{T}$ is associated with normalized port admittance matrix~$\M{y}$ and matrix~$\M{P}$ is associated with normalized matrix~$\M{U}$. In this way, the spherical waves act as ports which can be incorporated into a complete port characterization of the antenna similar to the antenna scattering matrix~\cite{1948_Montgomery_Principles_of_Microwave_Circuits,Hansen_SphericalNearFieldAntennaMeasurements} 
\begin{equation}
	\begin{bmatrix}
		\M{P}^{\trans}\M{Y}\M{P} & \M{P}^{\trans}\M{Y}\M{U}^{\trans}\\
		\M{U}\M{Y}\M{P} & \M{U}\M{Y}\M{U}^{\trans}\\
	\end{bmatrix}
	\begin{bmatrix}
		\M{v} \\
		\M{a} 		
	\end{bmatrix}
	=\begin{bmatrix}
		\M{i} \\
		-\M{f} 		
	\end{bmatrix}.
\label{eq:}
\end{equation}

The network characteristic modes~\cite{Mautz1973}
\begin{equation}
\M{y} \M{v}_n = \left( 1 -\J \lambda_n \right) \T{Re} \left\{\M{y} \right\}\M{v}_n
\label{eq:pp2}
\end{equation}
then directly translate to
\begin{equation}
\M{T} \M{f}_n = \left(1 - \J \lambda_n \right) \T{Re} \left\{\M{T} \right\}\M{f}_n,
\label{eq:ppx}
\end{equation}
which, combining~\eqref{eq:CM5} and~\eqref{eq:appUnit:3}, holds for loss-less scatterers.

\subsection{Substructure Modes in the Scattering Formulation}
\label{sec:disc:sub}

Substructure modes introduced in~\cite{Ethier_2012_substructure_TCM} have a potential to reduce the size of a problem and offer a good alternative to the network (port-mode) representation discussed in Section~\ref{sec:disc:unif}. Their evaluation is based on Schur complement~\cite{HornJohnson_MatrixAnalysis} which assumes an excitation field on the uncontrollable part of the structure which is solely due to currents on the controllable region. 

This assumption is, however, not met with the scattering approach~\eqref{eq:CM5}, since it was shown in~\eqref{eq:pp1} that the scattering formulation is formally similar to a port-mode representation. Such a description is in general incompatible with division into substructures. 

The scattering formulation is, therefore, not compatible with the substructure modes as defined in~\cite{Ethier_2012_substructure_TCM}, however, there might be an alternative approach devised in the future, which will preserve the essential properties of substructure modes (mainly the possibility to compress the size of the problem).

\section{Conclusion}
\label{sec:concl}

The focus of this work is the connection between scattering- and impedance-based formulations of characteristic modes, developing theoretical fundamentals which are subsequently applied in Part~II. While the impedance-based formulation is now the standard in the literature, it is demonstrated in this work that the scattering-based approach alleviates many of the known issues connected to characteristic modes decomposition.

The scattering-based approach studied in this work requires the computation of the transition matrix, a structure not typically available from \ac{MoM} numerical tools. However, computation of the transition matrix within a \ac{MoM} routine involves minimal computational overhead and only requires the calculation of basis function projections onto spherical harmonics. Also, the use of the transition matrix, as opposed to a \ac{MoM} impedance matrix, generalizes the calculation of characteristic modes to any computational method capable of producing a transition matrix, including finite difference time domain and finite element methods.  This lack of dependence on a chosen numerical method implies the consistent definition of characteristic modes for problems involving any lossless materials with no change to the underlying eigenvalue problem, greatly expanding the applicability of characteristic modes beyond \ac{MoM} formulations.

The examples in this work demonstrate several computational advantages arising from the use of scattering-based characteristic mode formulations. In some cases, significant improvements in numerical dynamics and computational cost are observed for single-frequency calculations.  

A few open problems remain in the use of scattering-based characteristic mode formulations.  Namely, the ability of characteristic modes to represent the behavior of systems excited by localized, near-field sources (\eg{}, discrete antenna ports) remains to be rigorously determined. Additionally, the issue of applying characteristic mode analysis to represent the behavior of disjoint or non-convex scatterers is elucidated by the scattering-based formulation.  The study of these fundamental questions from scattering-based perspectives will continue to steer the understanding and use of characteristic modes, regardless of the specific numerical tool or formulation adopted by individual practitioners.  

\appendices

\section{Normality of Transition Matrices and Unitarity of Scattering Matrices}
\label{app:Sunitary}
Using expansion~\eqref{eq:EHExpansion} and~\eqref{eq:EHIExpansion}, the net power exiting a surface~$S$ fully enclosing the scatter $\varOmega$
can be evaluated as~\cite{Kristensson_ScatteringBook}
\begin{equation}
\label{eq:appUnit:2}
P_S = \dfrac{1}{2} \T{Re} \oint \limits_S \left( \V{E} \times \V{H}^* \right) \cdot \T{d}\V{S} = \dfrac{1}{2} \left( \left| \M{f} \right|^2 + \T{Re} \left\{ \M{a}^\herm \M{f} \right\} \right),
\end{equation}
where the first term on the right-hand side represents the outward power flux, \ie{}, the scattered power in \eqref{eq:PradFF}, while the second term is minus the cycle mean power supplied by sources external to the scatterer.

For a lossless scatterer the net power~$P_S$ must vanish for any incident field described by vector~$\M{a}$. Substituting from~\eqref{eq:Tmat} 
there must be
\begin{equation}
    \label{eq:appUnit:3}
    \M{T}^\herm \M{T} = - \T{Re} \left\{ \M{T} \right\},
\end{equation}
where the real and imaginary part of a matrix are understood as
\begin{equation}
\begin{aligned}
    \T{Re} \left\{ \M{T} \right\} &= \dfrac{1}{2} \left( \M{T} + \M{T}^\herm \right), \\
    \T{Im} \left\{ \M{T} \right\} &= \dfrac{1}{2 \T{j}} \left( \M{T} - \M{T}^\herm \right).
\end{aligned}
\end{equation}
Relation~\eqref{eq:appUnit:3} leads to the following vanishing commutators
\begin{equation}
\Big[\T{Re} \left\{ \M{T} \right\}, \T{Im} \left\{ \M{T} \right\} \Big] = \M{0} \quad \text{and} \quad \Big[ \M{T} , \M{T}^\herm \Big] = \M{0}
\end{equation}
proving that the transition matrix~$\M{T}$ of a lossless scatterer is a normal matrix.  The spectral theorem~\cite[Chap.~2.5]{HornJohnson_MatrixAnalysis} states that a matrix is unitarily diagonalizable if and only if the matrix is normal, providing an unambiguous requirement for orthogonal eigenvectors.

Relation~\eqref{eq:appUnit:3} together with relation~\eqref{eq:Sdef} directly leads to
\begin{equation}
\Smat^\herm \Smat - \M{1} = \M{0}
\end{equation}
showing that the scattering matrix~$\Smat$ of a lossless scatterer is unitary~\cite[Chap.~4]{Collin_FoundationsForMicrowaveEngineering}. Lorentz reciprocity further implies~\cite[Chap.~4]{Collin_FoundationsForMicrowaveEngineering} that objects made of reciprocal materials have symmetric scattering matrices $\M{S}=\M{S}^{\T{T}}$ as well as transition matrices~$\M{T}=\M{T}^{\T{T}}$.

The unitarity of lossless, reciprocal scattering matrices implies unitary eigenvalues~$|s_n|=1$. Complex conjugation of the underlying eigenvalue problem~\eqref{eq:eigS},  multiplication with $\M{S}s_n$, and use of the symmetric nature of the scattering matrix gives     
\begin{equation}
	\M{S}\M{f}^{\ast}_n=s_n\M{f}_n^{\ast},
\end{equation} 
showing that~$\M{f}_n$ and $\M{f}^{\ast}_n$ are both eigenvectors of $\M{S}$. Hence, the phase of the eigenvector $\M{f}_n$ is arbitrary such that it may be chosen to be real valued.  

\section{Different Schemes For Conversion of Matrix~$\M{Z}$ Into Matrix~$\M{T}$.}
\label{App:IEformulations}

This appendix shows several of the most frequent forms of field integral equations, their method-of-moments formulations, and explicit forms of relations between the associated impedance and transition matrices. The formulation for penetrable bodies based on surface equivalence principle~\cite{ChangHarrington_AsurfaceFormulationForCharacteristicModesOfMaterialBodies} is not shown as it is detailed in Section~\ref{sec:CMfromT}. 

\subsection{$\Je$ -- EFIE}
\label{sec:JEFIE}
The electric field integral equation (EFIE) is the most common formulation of scattering from dielectric or highly electrically conductive obstacles. Employing the notation introduced in Section~\ref{sec:ElmagDescrip}, the integral equation reads
\begin{equation}
\label{eq:JEFIE:App}
     \rho \Je + \T{j} Z k \mathcal{L} \left\{ \Je \right\} = \Ei.
\end{equation}
In the case of volumetric formulation for dielectric obstacles, equality~\eqref{eq:JEFIE:App} is enforced throughout the obstacle, which is assumed to be described by the (possibly inhomogeneous) complex resistivity
\begin{equation}
    \rho =  -  \T{j} \dfrac{ Z}{k} \chi^{-1}
\end{equation}
with~$\chi$ being electric susceptibility. In the case of highly conducting obstacle, the equality~\eqref{eq:JEFIE:App} is enforced only at the surface of the obstacle and the term~$\rho \Je$ is substituted by~$Z_\T{s} \Je$ with~$Z_\T{s}$ representing a surface impedance~\cite{Jackson_ClassicalElectrodynamics,SenoirVolakis_ApproximativeBoundaryConditionsInEM} and with~$\Je$ being a surface current density.

A method-of-moments formulation in this case reads
\begin{equation}
    \label{eq:MoMJEFIE}
    \M{Z}^\T{ee} \Ie = \Ve
\end{equation}
and the transition matrix~$\M{T}$ can be evaluated as
\begin{equation}
\M{T} = - \Umat_1 \left( \M{Z}^\T{ee} \right)^{-1} \Umat_1^\T{T}
\end{equation}
and the radiation operator~$\M{R}_0$ reads
\begin{equation}
\label{eq:R0JEFIE}
    \M{R}_0 = \Umat_1^\herm \Umat_1
\end{equation}
and conversion from spherical expansion coefficients to current expansion coefficients of characteristic modes reads
\begin{equation}
    \label{eq:CM4JEFIE}
    \Ie_n = t_n^{-1} \left( \M{Z}^\T{ee} \right)^{-1} \Umat_1^\herm 
    \M{f}_n.
\end{equation}
The generalized eigenvalue problem for characteristic modes based on matrix~$\M{Z}^\T{ee}$ in this case leads to well-known relation~\cite{HarringtonMautz_TheoryOfCharacteristicModesForConductingBodies}
\begin{equation}
    \M{Z}^\T{ee} \Ie_n = \left(1 + \T{j}\lambda_n \right) \M{R}_0 \Ie_n,
    \label{eq:app:EFIEGEP}
\end{equation}
where in loss-less cases~$\T{Re} \left\{ \M{Z}^\T{ee} \right\} = \M{R}_0$.

\subsection{$\Je$ -- MFIE}
\label{sec:JMFIE}
Another common formulation of scattering from closed perfectly electrically conducting objects is the magnetic field integral equation (MFIE), which reads
\begin{equation}
\label{eq:JMFIE:App}
     - \dfrac{1}{2} \hat{\V{n}} \times \Je - \mathcal{K} \left\{ \Je \right\}  = \Hi.
\end{equation}
The equation is imposed over the closed conducting surface $\varOmega$ with outward-directed normal~$\hat{\V{n}}$. Using the notation from~Section~\ref{sec:CMfromT}, the method-of-moments formulation in this case reads
\begin{equation}
    \label{eq:MoMJMFIE}
    \M{Z}^\T{me} \Ie = \Vm
\end{equation}
and the~$\M{Z} \to \M{T}$ conversion formula reads
\begin{equation}
    \label{eq:MoMJMFIET}
   \M{T} =  \Umat_1 \left( \T{j}\M{Z}^\T{me} \right)^{-1} \overline{\Umat}_1^\T{T}.
\end{equation}
Characteristic currents are in this case evaluated as
\begin{equation}
    \Ie_n = - t_n^{-1} \left( \T{j}\M{Z}^\T{me} \right)^{-1} \overline{\Umat}_1^\trans \M{f}_n,
\end{equation}
which can also be transformed into a generalized eigenvalue problem
\begin{equation}
    -\T{j}\M{Z}^\T{me} \Ie_n = \left(1 + \T{j}\lambda_n \right) \overline{\Umat}_1^\trans \Umat_1 \Ie_n,
    \label{eq:app:MFIE:GEP}
\end{equation}
which has a form proposed in~\cite{DaiLiuGanChew_CFIEforCMs} assuming~$\T{Im} \left\{ \M{Z}^\T{me} \right\} = \overline{\Umat}_1^\trans \Umat_1$, which holds for lossless scatterers.

Problem~\eqref{eq:app:MFIE:GEP} resembles~\eqref{eq:app:EFIEGEP} but there are notable differences. The radiation operator~$\M{R}_0$ is still given by~\eqref{eq:R0JEFIE} and is not equal to the matrix appearing on the right-hand side of~\eqref{eq:app:MFIE:GEP}. This invalidates the diagonalization of the system matrix~$\M{Z}^\T{me}$ with the eigencurrents~$\Ie_n$. Instead, and similarly to the left and right eigenvectors in~\cite{YlaOijala_GeneralizedTCM2019}, a dual current
\begin{equation}
    \overline{\M{I}}^{\T{e}}_n = - t_n^{-1} \left( \T{j}\M{Z}^\T{me} \right)^{-\T{T}} \Umat_1^\trans \M{f}_n,
\end{equation}
is introduced by exchanging TM spherical waves and TE spherical waves which is motivated by the off diagonal position of matrix~$\M{Z}^\T{me}$ in the impedance matrix~\eqref{eq:ZIVSurfEqv}. Using that the matrix~$\M{T}$ is symmetric (reciprocity), it can be seen that
\begin{equation}
    \M{f}_n = - \overline{\Umat}_1 \overline{\M{I}}^{\T{e}}_n
\end{equation}
and hence
\begin{multline}
    - \big(\overline{\M{I}}^{\T{e}}_m\big)^\herm  \T{j}\M{Z}^\T{me} \Ie_n = (1 + \T{j}\lambda_n ) \big(\overline{\M{I}}^{\T{e}}_m\big)^\herm \overline{\Umat}_1^\trans \Umat_1 \Ie_n \\
    = (1 + \T{j}\lambda_n)\M{f}_m^{\herm}\M{f}_n 
    = (1 + \T{j}\lambda_n)\delta_{mn} 
\end{multline}
which is consistent with~\eqref{eq:IZI} if the magnetic current is identified with the dual current.

\subsection{$\Je$ -- CFIE}
In order to avoid internal resonance problems~\cite{1978_Harrington_TechRep,ChewTongHu_IntegralEquationMethodsForElectromagneticAndElasticWaves}, a linear combination of the EFIE~\eqref{eq:JEFIE:App} and MFIE~\eqref{eq:MoMJMFIE} formulations
\begin{equation}
    \alpha \T{EFIE} + Z \left( 1 - \alpha  \right) \hat{\V{n}} \times \T{MFIE}
\end{equation}
with~$\alpha \in \left(0,1 \right)$ is commonly imposed over the surface of a closed perfectly conducting obstacle to solve the corresponding scattering scenario. The resulting equation is called the combined field integral equation (CFIE). The CFIE involves only electric currents represented by $\Ie$,
\begin{equation}
    \label{eq:MoMJCFIE}
    \M{Z}^\T{ee} \Ie = \alpha \Ve + Z \left( 1 - \alpha  \right) \Vm
\end{equation}
and its part coming from EFIE is identical to~Appendix~\ref{sec:JEFIE}. The part $\hat{\V{n}} \times \T{MFIE}$ however differs from MFIE from Appendix~\ref{sec:JMFIE}. In particular, the excitation vector corresponding to incident magnetic field now reads
\begin{equation}
    \label{eq:nW}
    V^\T{m}_n = \int \limits_{\srcRegion} \Big( \hat{\V{n}} \left( \V{r} \right) \times \Hi\left( \V{r} \right) \Big) \cdot \V{\psi}_n \left( \V{r} \right) \T{d}S,
\end{equation}
or equivalently,
\begin{equation}
    \T{j} \Vm = 
    - \left( \overline{\Umat}_1^\T{n} \right)^\T{T} \M{a},
\end{equation}
with
\begin{equation}
    \label{eq:nSmatrix}
    \overline{\Umat}_1^\T{n} = \dfrac{k}{\sqrt{Z}} \left[ \, \int \limits_{\srcRegion} \Big( \hat{\V{n}} \left( \V{r} \right) \times \M{u}_{\overline{\alpha}}^{\left(1\right)} \left(k \V{r} \right) \Big) \cdot \V{\psi}_n \left( \V{r} \right) \T{d}S \right].
\end{equation}

An algebraic relation between $\M{Z}$ and $\M{T}$ is given by
\begin{equation}
\M{T} = -\Umat_1 \left( \M{Z}^\T{ee} \right)^{-1} \left( \alpha \Umat_1 + \T{j} Z\left( 1 - \alpha  \right)  
 \overline{\Umat}_1^\T{n} \right)^\T{T}.  
\end{equation}
Characteristic currents are in this case evaluated as
\begin{equation}
    \Ie_n = t_n^{-1} \left( \M{Z}^\T{ee} \right)^{-1} \left( \alpha \Umat_1 + \T{j} Z \left( 1 - \alpha  \right) \overline{\Umat}_1^\T{n} \right)^\T{T} \M{f}_n
\end{equation}
which can also be transformed into a generalized eigenvalue problem
\begin{equation}
    \M{Z}^\T{ee} \Ie_n = \left(1 + \T{j}\lambda_n \right) \left( \alpha \Umat_1 + \T{j} Z \left( 1 - \alpha  \right) \overline{\Umat}_1^\T{n} \right)^\T{T} \Umat_1 \Ie_n.
\end{equation}

\subsection{PMCHWT}
The PMCHWT formulation for a homogeneous dielectric object can be considered as combining EFIE and MFIE in the interior and exterior regions
\begin{equation}
    \M{Z}_{\T{i}}\M{I}
    =
    \begin{bmatrix}
     \M{Z}_{\T{i}}^{\T{ee}} & \J\M{Z}_{\T{i}}^{\T{em}}\\
     \J\M{Z}_{\T{i}}^{\T{me}} & \M{Z}_{\T{i}}^{\T{mm}}
    \end{bmatrix}
    \begin{bmatrix}
     \M{I}^{\T{e}} \\ \J\M{I}^{\T{m}} 
    \end{bmatrix}
    = \begin{bmatrix}
     \M{0} \\ \M{0} 
    \end{bmatrix}
    \label{eq:PMCHWT1}
\end{equation}
and
\begin{equation}
    \M{Z}_{\T{o}}\M{I}
    =
    \begin{bmatrix}
     \M{Z}_{\T{o}}^{\T{ee}} & \J\M{Z}_{\T{o}}^{\T{em}}\\
     \J\M{Z}_{\T{o}}^{\T{me}} & \M{Z}_{\T{o}}^{\T{mm}}
    \end{bmatrix}
    \begin{bmatrix}
     \M{I}^{\T{e}} \\ \J\M{I}^{\T{m}} 
    \end{bmatrix}
    = \begin{bmatrix}
     \M{V}^{\T{e}} \\ \J\M{V}^{\T{m}}
    \end{bmatrix}
    \label{eq:PMCHWT2}
\end{equation}
where $\M{Z}_{\T{i/o}}^{mn}$ is constructed by using the permittivity in the inner (i) and outer (o) region. The diagonal and off diagonal terms are determined by the EFIE ($\mathcal{L}$) and MFIE ($\mathcal{K}$ and $\hat{\V{n}}\times$, cf~\eqref{eq:JMFIE:App}) operators, respectively~\cite{Jin_TheoryAndComputationOfElectromagneticFields,ChewTongHu_IntegralEquationMethodsForElectromagneticAndElasticWaves}. Adding the matrices in~\eqref{eq:PMCHWT1} and~\eqref{eq:PMCHWT2} such that $\hat{\V{n}}\times$ is eliminated defines the PMCHWT formulation. Here, it is noted that the radiation operator~\eqref{eq:R0} is
\begin{equation}
    \M{R}_0 = \Umat^\herm \Umat
    = \T{Re} \left\{\M{Z}_\T{o} \right\}
\end{equation}
making~\eqref{eq:CMI} similar to the exterior formulation in~\cite{YlaOijala_PMCHWTBasedCharacteristicModeFormulationsforMaterialBodies}.  

\bibliographystyle{IEEEtran}
\bibliography{references,extraBib}


\begin{IEEEbiography}[{\includegraphics[width=1in,height=1.25in,clip,keepaspectratio]{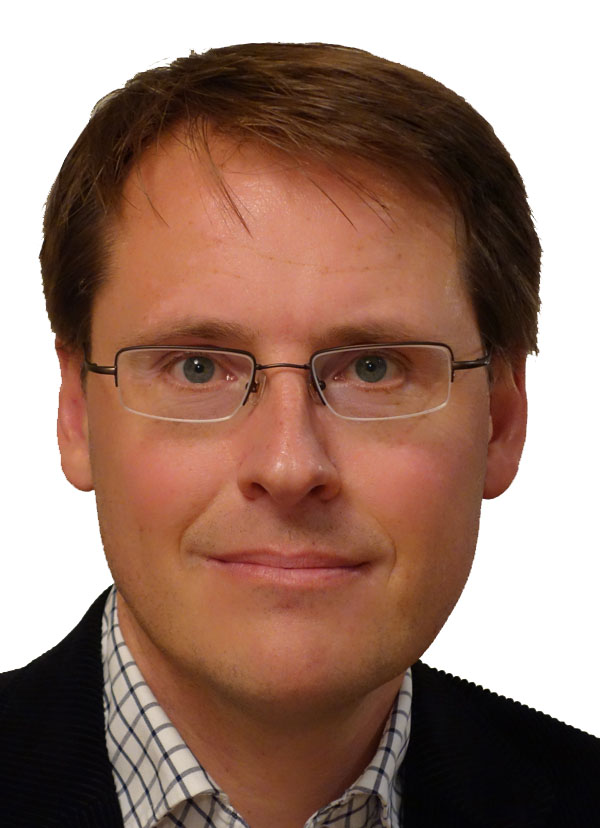}}]{Mats Gustafsson}
received the M.Sc. degree in Engineering Physics 1994, the Ph.D. degree in Electromagnetic Theory 2000, was appointed Docent 2005, and Professor of Electromagnetic Theory 2011, all from Lund University, Sweden.

He co-founded the company Phase holographic imaging AB in 2004. His research interests are in scattering and antenna theory and inverse scattering and imaging. He has written over 100 peer reviewed journal papers and over 100 conference papers. Prof. Gustafsson received the IEEE Schelkunoff Transactions Prize Paper Award 2010, the IEEE Uslenghi Letters Prize Paper Award 2019, and best paper awards at EuCAP 2007 and 2013. He served as an IEEE AP-S Distinguished Lecturer for 2013-15.
\end{IEEEbiography}

\begin{IEEEbiography}[{\includegraphics[width=1in,height=1.25in,clip,keepaspectratio]{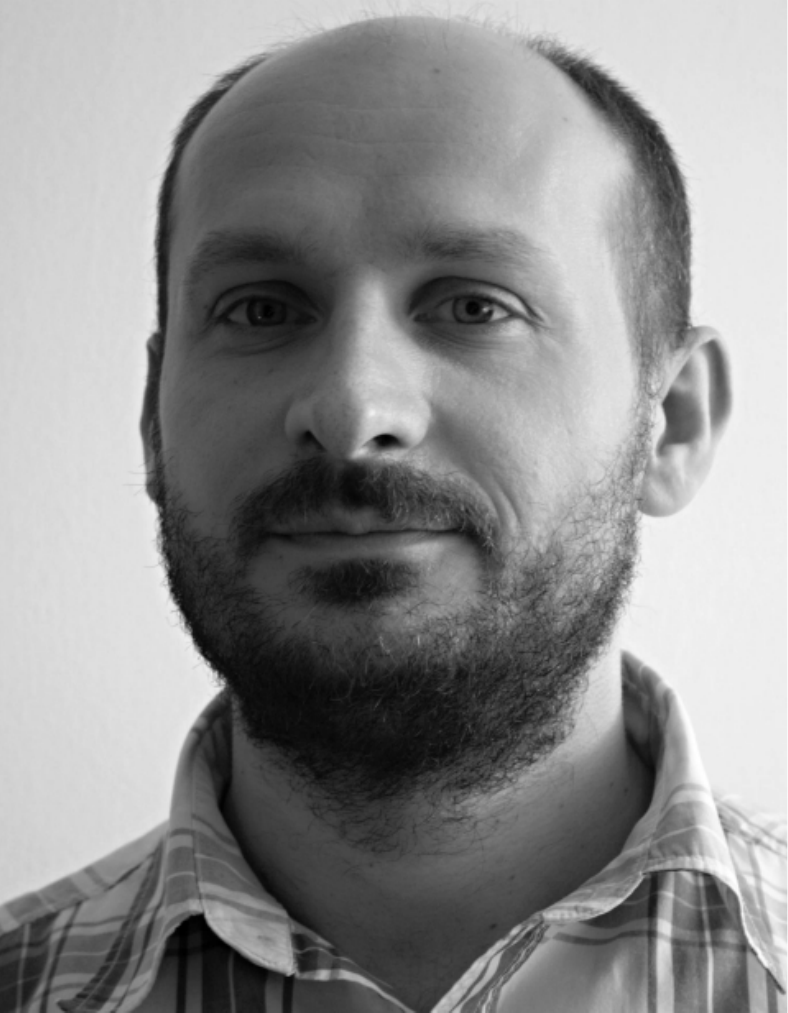}}]{Lukas Jelinek}
received his Ph.D. degree from the Czech Technical University in Prague, Czech Republic, in 2006. In 2015 he was appointed Associate Professor at the Department of Electromagnetic Field at the same university.

His research interests include wave propagation in complex media, electromagnetic field theory, metamaterials, numerical techniques, and optimization.
\end{IEEEbiography}

\begin{IEEEbiography}[{\includegraphics[width=1in,height=1.25in,clip,keepaspectratio]{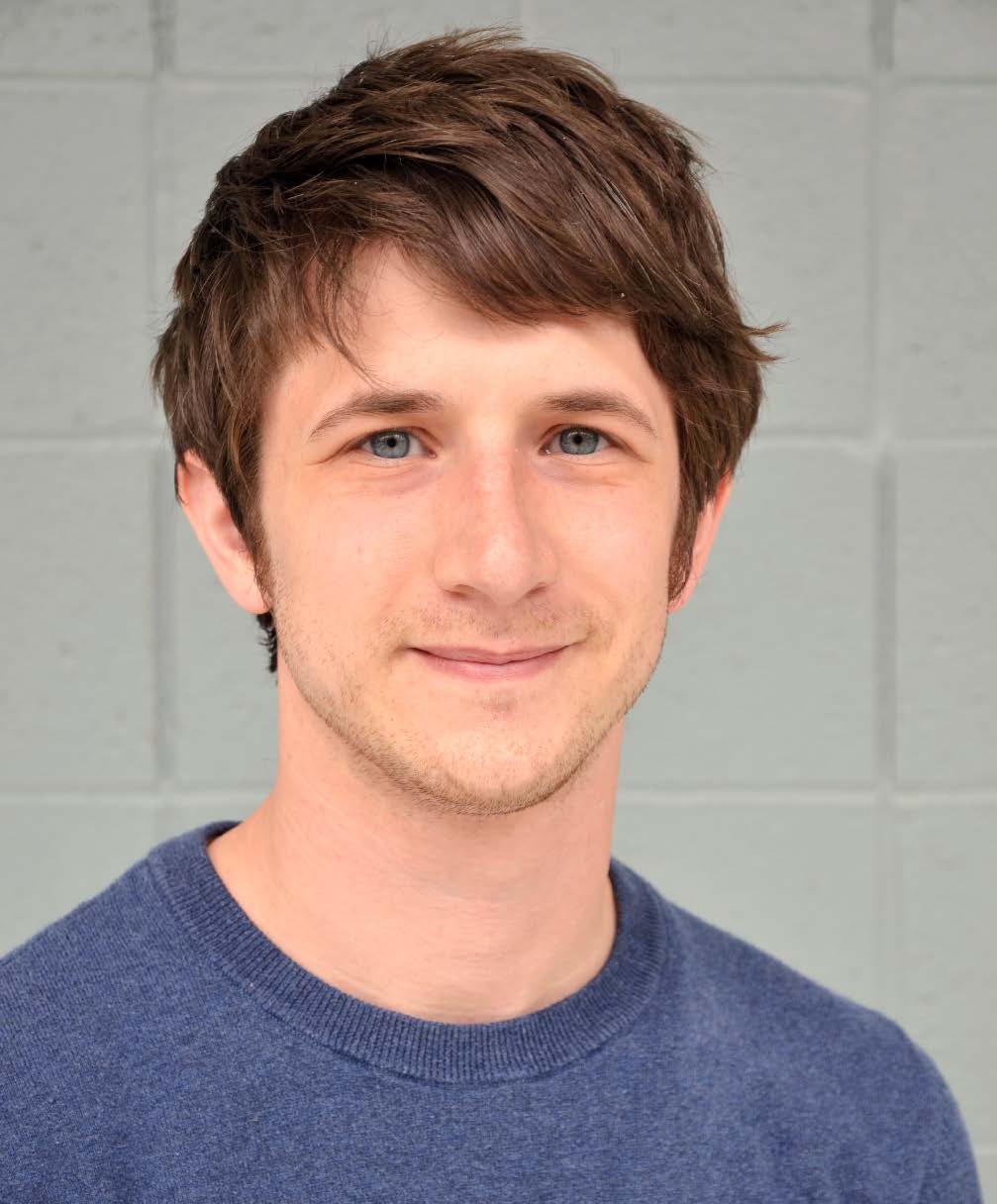}}]{Kurt Schab}(M'16)
Kurt Schab is an Assistant Professor of Electrical Engineering at Santa Clara University, Santa Clara, CA USA. He received the B.S. degree in electrical engineering and physics from Portland State University in 2011, and the M.S. and Ph.D. degrees in electrical engineering from the University of Illinois at Urbana-Champaign in 2013 and 2016, respectively.  From 2016 to 2018 he was an Intelligence Community Postdoctoral Research Scholar at North Carolina State University in Raleigh, North Carolina.  His research focuses on the intersection of numerical methods, electromagnetic theory, and antenna design.  
\end{IEEEbiography}

\begin{IEEEbiography}[{\includegraphics[width=1in,height=1.25in,clip,keepaspectratio]{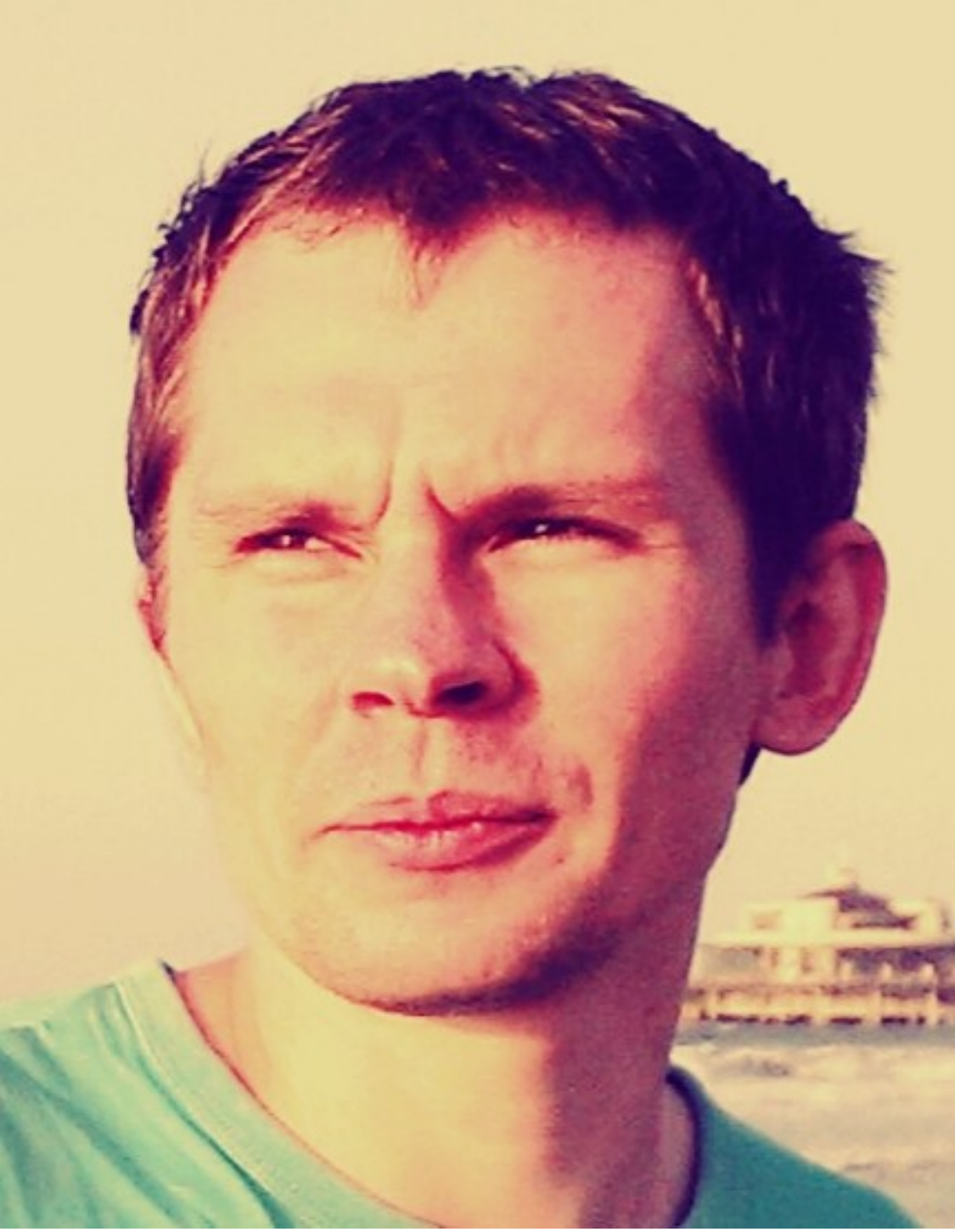}}]{Miloslav Capek}
(M'14, SM'17) received the M.Sc. degree in Electrical Engineering 2009, the Ph.D. degree in 2014, and was appointed Associate Professor in 2017, all from the Czech Technical University in Prague, Czech Republic.
	
He leads the development of the AToM (Antenna Toolbox for Matlab) package. His research interests are in the area of electromagnetic theory, electrically small antennas, numerical techniques, and optimization. He authored or co-authored over 100~journal and conference papers.

Dr. Capek is member of Radioengineering Society and Associate Editor of IET Microwaves, Antennas \& Propagation.
\end{IEEEbiography}

\end{document}